# EGFR mutation prediction using $F^{18}$-FDG PET-CT based radiomics features in non-small cell lung cancer

H. Henríquez[1,2], D. Fuentes[2], F. Suarez[3], P. Gonzalez[4]

[1] Clínica Santa María, Servicio de Radiología, Santiago, Chile.
[2] Universidad de los Andes, Facultad de Medicina, Santiago, Chile.
[3] Clínica Santa María, Unidad de Cirugía de Tórax, Santiago, Chile.
[4] Clínica Santa María, Unidad de Medicina Nuclear, Santiago, Chile.

*Abstract*— Lung cancer is the leading cause of cancer death in the world. Accurate determination of the EGFR (epidermal growth factor receptor) mutation status is highly relevant for the proper treatment of this patients. Purpose: The aim of this study was to predict the mutational status of the EGFR in non-small cell lung cancer patients using radiomics features extracted from PET-CT images. Methods: Retrospective study that involve 34 patients with lung cancer confirmed by histology and EGFR status mutation assessment. A total of 2.205 radiomics features were extracted from manual segmentation of the PET-CT images using pyradiomics library. Both computed tomography and positron emission tomography images were used. All images were acquired with intravenous iodinated contrast and $F^{18}$-FDG. Preprocessing includes resampling, normalization, and discretization of the pixel intensity. Three methods were used for the feature selection process: backward selection (set 1), forward selection (set 2), and feature importance analysis of random forest model (set 3). Nine machine learning methods were used for radiomics model building. Results: 35.2% of patients had EGFR mutation, without significant differences in age, gender, tumor size and SUVmax. After the feature selection process 6, 7 and 17 radiomics features were selected, respectively in each group. The best performances were obtained by Ridge Regression in set 1: AUC of 0.826 (95% CI, 0.811 – 0.839), Random Forest in set 2: AUC of 0.823 (95% CI, 0.808 – 0.838) and Neural Network in set 3: AUC of 0.821 (95% CI, 0.808 – 0.835). Conclusion: The radiomics features analysis has the potential of predicting clinically relevant mutations in lung cancer patients through a non-invasive methodology.

## I. Introduction

Lung cancer is the leading cause of cancer death in the world [1]. In patients with epidermal growth factor receptor (EGFR) mutated tumors, the inhibitors of tyrosine kinase receptor (TKIs) have been shown to significantly improve overall survival [2]. Accurate determination of the EGFR mutation status is highly relevant for the proper treatment of this patients. The genetic test to determine the mutation of the EGFR involves an invasive procedure in obtaining the tumor tissue sample with risk of complications. In addition, its availability is limited in some regions of the world.

Medical images are the most used diagnostic modality in cancer patients, provides information about diagnosis and prognosis. Advances in technology have allowed for improved image resolution, standardization of protocols, and global availability [3].

The radiomics features analysis is a non-invasive methodology that converts imaging into high dimensional data, through automatic feature extraction, and has shown a good correlation with histological subtypes of tumors and genetic status in several pathologies, including lung cancer and squamous cell carcinomas of the head and neck [4,5].

The aim of this study was to predict the mutational status of the EGFR using radiomics features extracted from positron emission tomography – computed tomography (PET-CT) images in non-small cell lung cancer patients.

## II. Materials and methods

### A. Patients:

This study was approved by the Ethics Committee of Clínica Santa María and a waiver for the informed consent was obtained.

Retrospective study of patients with lung cancer treated with surgery from 2015 to 2020, confirmed by histology.

Inclusion criteria: 1) patients older than 18 years, 2) preoperative $F^{18}$-FDG PET-CT with intravenous contrast, and 3) available biopsy with EGFR mutation study.

Exclusion criteria: 1) patients with images that had artifacts of any kind that would prevent adequate lesion segmentation, 2) non-measurable lesions, and 3) studies without intravenous contrast.

The clinical data collected included: age, gender, stage, tumor histology, EGFR mutation status and SUVmax (maximum standard uptake value).

*B. Image acquisition:*

All images were acquired with 16-channel Philips TruFlight Select PET-CT equipment. Acquisition parameters: 120 kVp, 100 – 210 mAs and 512 x 512 matrix. The acquisition protocol considered positron emission tomography (PET) time of light (TOF). Computed tomography (CT) images were acquired with intravenous contrast in a venous phase with slice thickness of 3 mm and an exclusive late phase of the chest with a slice thickness of 1 mm. All studies were performed with $F^{18}$- fluordoxiglucose ($F^{18}$-FDG).

*C. Tumor segmentation:*

Two groups of images were segmented in all patients: late-phase chest CT images and PET images (Figure 1).

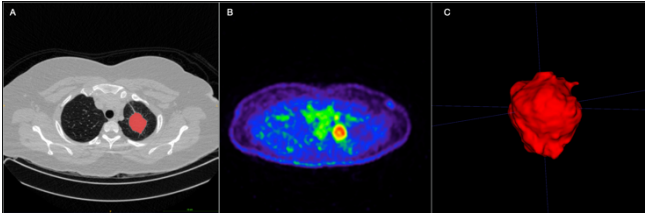

Fig. 1. A: Chest CT image with tumor segmentation in the left upper lobe. B: PET image showing intense $F^{18}$- FDG uptake. C: Tumor volumetric segmentation.

Segmentations were performed by two operators using volumetric technique in ITK-Snap software (H.H., radiologist with 10 years of experience in oncological imaging and D.F., third year radiology resident) [6]. All segmentations were evaluated by the two operators and in cases of doubt about the boundaries of the lesion, the decision was made by the most experienced radiologist. The thresholding and pixel growing tool were used in most tumors, with manual correction of tumor boundaries in selected cases.

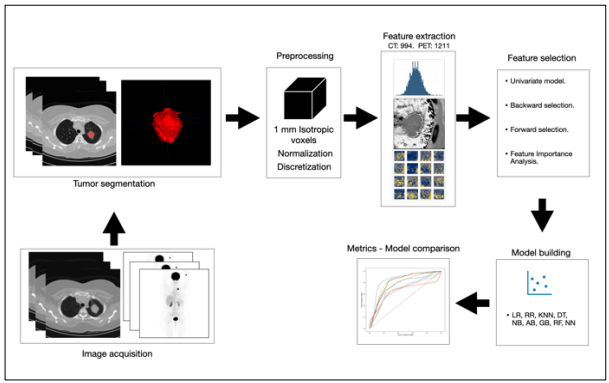

Fig. 2. Workflow summary.

*D. Preprocessing and radiomics feature extraction:*

The preprocessing and feature extraction was performed with pyradiomics library [7]. Preprocessing included pixel intensity normalization, discretization (bin = 5) and resampling of the images to 1 mm isotropic voxels.

Filters were applied to the extracted features (wavelet, log, square, squareroot, exponential and logarithm).

Table 1. Patients' characteristics: AdenoCa: Adenocarcinoma; SCCa: Squamous Cell Carcinoma.
Mean ± std; count (%).

| Features | EGFR (+) | EGFR (-) | p-value |
|---|---|---|---|
| Age | 61,0 ± 14,8 | 67,7 ± 10,2 | 0,061 |
| Gender: | | | 0,472 |
|     female | 7 (58,3) | 10 (45,5) | |
|     male | 5 (41,7) | 12 (54,5) | |
| Stage: | | | 0,194 |
|     I | 0 | 3 (13,6) | |
|     II | 0 | 3 (13,6) | |
|     III | 2 (16,7) | 3 (13,6) | |
|     IV | 10 (83,3) | 13 (59,1) | |
| Size | 48,9 ±25 | 39,6 ± 20,9 | 0,256 |
| SUVmax | 11,3 ± 5,2 | 13,2 ± 8,5 | 0,614 |
| Histology: | | | 0,535 |
|     AdenoCa | 12 (100) | 21 (95,5) | |
|     SCCa | 0 | 1 (4,5) | |
|     Other | 0 | | |
| EGFR mutation: | | | |
|   Exon19 | 7 (58,3) | - | - |
|   Exon20 | 1 (8,3) | - | - |
|   Exon21 | 4 (33,3) | - | - |

*E. Feature Selection:*

For the dimensionality reduction process, only the features that reached statistical significance (p-value < 0.05) in the univariate analysis were included. Subsequently, three sets of features were created by different methods: backward selection (set 1), forward selection (set 2), and feature importance analysis of Random Forest model (set 3). For all methods, 5-fold cross validation was performed. For the feature importance analysis, a hyperparameter search was first performed by gridsearch with all features, and 5-fold cross validation was repeated 100 times. Features with feature importance greater than 0 and which were repeated with greater frequency were selected (0.9 percentil).

## F. Model building:

The Machine Learning methods used were loaded from Sklearn and Keras libraries. The hyperparameter search was performed with gridsearch and the performance metrics were calculated with 100 repetitions of 5-fold cross-validation. Nine Machine Learning models were fitted to the 3 datasets: Logistic Regression (LR), Ridge Regression (RR), K-Nearest Neighbor (KNN), Naive Bayes (NB), Decision Tree (DT), Random Forest (RF), AdaBoost (AB), Gradient Boosting (GB) and Neural Network (NN).

Table 2. Radiomics features selected for each set (set 1: backward selection, set 2: forward selection and set 3: feature importance analysis).

| Set | Selected features |
|---|---|
| Set 1 | CT-square firstorder Skewness<br>PET-original glszm ZoneVariance<br>PET-wavelet-LHLglszmSmallAreaHighGrayLevel<br>PET-wavelet-HLL glcm Idn<br>PET-wavelet-HLL glcm InverseVariance<br>PET-wavelet-HLH glcm Correlation |
| Set 2 | CT-logarithm firstorder Skewness<br>CT-wavelet-HLH glszm SmallAreaEmphasis<br>CT-wavelet-HHH glrlm GrayLevelVariance<br>CT-wavelet-HLH glrlm ShortRunLowGrayLevelEmphasis<br>PET-wavelet-HHL firstorder Median<br>PET-wavelet-LHL glrlm ShortRunHighGrayLevEmphasis<br>PET-wavelet-HLL glcm Idmn |
| Set 3 | CT-square firstorder Skewness<br>PET-wavelet-HLL glcm InverseVariance<br>PET-wavelet-HLL glcm JointEntropy<br>PET-wavelet-HLL glcm SumEntropy<br>PET-wavelet-HLL gldm DependenceEntropy<br>PET-wavelet-HLL ngtdm Contrast<br>PET-wavelet-HLL glcm DifferenceAverage<br>PET-wavelet-HLL glcm Id<br>PET-wavelet-HLL glcm Idn<br>PET-wavelet-HLL glcm JointEnergy<br>PET-wavelet-HLL glcm Contrast<br>PET-wavelet-HLL glcm DifferenceVariance<br>PET-wavelet-HLL glcm Idmn<br>PET-wavelet-HLL gldm LargeDependLowGrayLevEmph<br>PET-wavelet-HLL glcm DifferenceEntropy<br>PET-wavelet-HLL glcm Idm<br>PET-wavelet-HLL ngtdm Complexity |

## G. Statistical Analysis:

Mann-Whitney U test and Chi-Square test were used for continuous and categorical variables, respectively.

Features with two-sided p-values < 0.05 were considered to have statistical significance. Accuracy, recall, precision, f1-score, roc-curves, and area under the curve (AUC) were calculated in all models. The threshold was defined by maximizing the f1-score.

## III. RESULTS

### A. Clinical characteristics:

A total of 130 patients with lung cancer treated with surgery from 2015 to 2020 were identified. We excluded 96 patients (71 because they did not have EGFR mutational study; 23 because lack of intravenous contrast studies, and 2 patients for having non-measurables lesions).

The patients who met the inclusion criteria were 34 (50% females), 12 of them (35,2%) had EGFR mutation.

The statistics of clinical characteristics of patients were shown in Table 1.

There were no significant differences in age, gender, tumor size, SUVmax or histological subtypes.

The most frequent histological subtype was adenocarcinoma (97,1%) followed by squamous cell carcinoma (2,9%). The most frequent EGFR mutation was exon 19 deletion (58,3%), followed by exon 21 deletion (33,3%).

### B. Radiomics features selection:

2.205 radiomics features were respectively extracted from PET and CT images (994 features from CT and 1.211 features from PET images).

151 features were considered statistically significant in the univariate analysis (32 from CT images and 119 from PET images).

Using a backward and forward selection methods, sets of 6 and 7 features were selected, respectively. 17 features were selected from Random Forest feature importance analysis (Table 2). In the case of obtaining redundant features with two or more different filters, it was decided to eliminate the variable that had the lowest AUC in the univariate model.

### C. Cross-validation results:

The three best cross validation AUC on the test set were: Set 1: Ridge Regression 0.826 (95% CI, 0.811 – 0.839), Random Forest 0.823 (95% CI, 0.809 – 0.837) and Gradient Boost 0.801 (95% CI, 0.786 – 0.816); Set 2: Random Forest 0.823 (95% CI, 0.808 – 0.838), GradientBoost 0.776 (95% CI, 0.761 – 0.791) and Ada Boost 0.735 (95% CI, 0.718 – 0.751); Set 3: Neural Network 0.821 (95% CI, 0.808 – 0.835), Ridge Regression 0.814 (95% CI, 0.801 – 0.827) and Naive Bayes 0.797 (95% CI, 0.783 – 0.811) (Figure 3).

Complete results of cross-validation are in Table 3.

| SET | Model | TRAIN | | | | | TEST | | | | |
|---|---|---|---|---|---|---|---|---|---|---|---|
| | | Accuracy | F1-Score | Precision | Recall | AUC | Accuracy | F1-Score | Precision | Recall | AUC |
| SET 1 | LR | 0.837 ± 0.060 | 0.836 ± 0.059 | 0.841 ± 0.059 | 0.833 ± 0.070 | 0.837 ± 0.060 | 0.760 ± 0.158 | 0.629 ± 0.282 | 0.631 ± 0.321 | 0.700 ± 0.329 | 0.752 ± 0.180 |
| | RF | 0.911 ± 0.031 | 0.871 ± 0.046 | 0.888 ± 0.064 | 0.859 ± 0.057 | 0.900 ± 0.035 | 0.843 ± 0.124 | 0.708 ± 0.287 | 0.757 ± 0.327 | 0.719 ± 0.318 | 0.823 ± 0.158 |
| | KNN | 0.769 ± 0.040 | 0.544 ± 0.130 | 0.872 ± 0.164 | 0.419 ± 0.147 | 0.686 ± 0.059 | 0.647 ± 0.157 | 0.283 ± 0.297 | 0.396 ± 0.430 | 0.268 ± 0.316 | 0.582 ± 0.158 |
| | DT | 0.900 ± 0.047 | 0.871 ± 0.058 | 0.807 ± 0.091 | 0.959 ± 0.068 | 0.913 ± 0.042 | 0.738 ± 0.170 | 0.604 ± 0.299 | 0.593 ± 0.333 | 0.699 ± 0.356 | 0.737 ± 0.189 |
| | NB | 0.660 ± 0.064 | 0.527 ± 0.119 | 0.835 ± 0.076 | 0.391 ± 0.117 | 0.660 ± 0.064 | 0.736 ± 0.162 | 0.444 ± 0.337 | 0.572 ± 0.433 | 0.408 ± 0.344 | 0.662 ± 0.181 |
| | AB | 0.972 ± 0.036 | 0.962 ± 0.049 | 0.949 ± 0.071 | 0.979 ± 0.049 | 0.974 ± 0.036 | 0.763 ± 0.158 | 0.597 ± 0.305 | 0.655 ± 0.350 | 0.620 ± 0.352 | 0.739 ± 0.185 |
| | GB | 0.991 ± 0.014 | 0.991 ± 0.014 | 0.984 ± 0.025 | 0.998 ± 0.010 | 0.991 ± 0.014 | 0.814 ± 0.147 | 0.663 ± 0.311 | 0.709 ± 0.350 | 0.691 ± 0.349 | 0.801 ± 0.171 |
| | RR | 0.839 ± 0.048 | 0.836 ± 0.052 | 0.848 ± 0.047 | 0.828 ± 0.076 | 0.839 ± 0.048 | 0.830 ± 0.139 | 0.737 ± 0.254 | 0.737 ± 0.291 | 0.803 ± 0.285 | 0.826 ± 0.159 |
| | NN | 0.841 ± 0.051 | 0.838 ± 0.054 | 0.853 ± 0.056 | 0.828 ± 0.078 | 0.841 ± 0.051 | 0.789 ± 0.146 | 0.675 ± 0.261 | 0.685 ± 0.303 | 0.743 ± 0.309 | 0.792 ± 0.156 |
| SET 2 | LR | 0.837 ± 0.065 | 0.838 ± 0.066 | 0.832 ± 0.064 | 0.847 ± 0.080 | 0.837 ± 0.065 | 0.697 ± 0.161 | 0.535 ± 0.288 | 0.541 ± 0.325 | 0.617 ± 0.358 | 0.684 ± 0.189 |
| | RF | 0.914 ± 0.031 | 0.881 ± 0.041 | 0.862 ± 0.062 | 0.904 ± 0.049 | 0.912 ± 0.032 | 0.829 ± 0.147 | 0.718 ± 0.276 | 0.742 ± 0.310 | 0.758 ± 0.313 | 0.823 ± 0.170 |
| | KNN | 0.778 ± 0.051 | 0.782 ± 0.054 | 0.768 ± 0.053 | 0.803 ± 0.088 | 0.778 ± 0.051 | 0.717 ± 0.159 | 0.596 ± 0.270 | 0.575 ± 0.303 | 0.693 ± 0.328 | 0.715 ± 0.190 |
| | DT | 0.750 ± 0.088 | 0.666 ± 0.148 | 0.636 ± 0.184 | 0.746 ± 0.202 | 0.751 ± 0.083 | 0.661 ± 0.204 | 0.501 ± 0.311 | 0.508 ± 0.354 | 0.591 ± 0.379 | 0.655 ± 0.205 |
| | NB | 0.840 ± 0.060 | 0.838 ± 0.065 | 0.842 ± 0.054 | 0.839 ± 0.094 | 0.840 ± 0.060 | 0.693 ± 0.176 | 0.528 ± 0.294 | 0.562 ± 0.351 | 0.586 ± 0.356 | 0.677 ± 0.196 |
| | AB | 0.984 ± 0.024 | 0.984 ± 0.024 | 0.980 ± 0.033 | 0.988 ± 0.030 | 0.984 ± 0.024 | 0.748 ± 0.161 | 0.591 ± 0.300 | 0.617 ± 0.344 | 0.652 ± 0.360 | 0.735 ± 0.187 |
| | GB | 0.997 ± 0.009 | 0.997 ± 0.009 | 0.997 ± 0.013 | 0.997 ± 0.013 | 0.997 ± 0.009 | 0.773 ± 0.160 | 0.638 ± 0.285 | 0.657 ± 0.327 | 0.703 ± 0.340 | 0.776 ± 0.172 |
| | RR | 0.863 ± 0.051 | 0.866 ± 0.050 | 0.846 ± 0.054 | 0.891 ± 0.067 | 0.863 ± 0.051 | 0.721 ± 0.155 | 0.577 ± 0.277 | 0.576 ± 0.313 | 0.669 ± 0.347 | 0.718 ± 0.176 |
| | NN | 0.942 ± 0.032 | 0.944 ± 0.032 | 0.917 ± 0.041 | 0.974 ± 0.039 | 0.942 ± 0.032 | 0.695 ± 0.162 | 0.513 ± 0.290 | 0.562 ± 0.353 | 0.562 ± 0.356 | 0.676 ± 0.186 |
| SET 3 | LR | 0.862 ± 0.052 | 0.865 ± 0.051 | 0.850 ± 0.058 | 0.883 ± 0.058 | 0.862 ± 0.052 | 0.762 ± 0.150 | 0.633 ± 0.276 | 0.636 ± 0.319 | 0.719 ± 0.332 | 0.763 ± 0.168 |
| | RF | 0.908 ± 0.028 | 0.863 ± 0.042 | 0.899 ± 0.055 | 0.833 ± 0.058 | 0.891 ± 0.032 | 0.817 ± 0.126 | 0.653 ± 0.300 | 0.721 ± 0.346 | 0.659 ± 0.337 | 0.791 ± 0.160 |
| | KNN | 0.898 ± 0.041 | 0.894 ± 0.045 | 0.928 ± 0.052 | 0.867 ± 0.070 | 0.898 ± 0.041 | 0.773 ± 0.149 | 0.657 ± 0.267 | 0.650 ± 0.308 | 0.755 ± 0.320 | 0.779 ± 0.165 |
| | DT | 0.947 ± 0.030 | 0.930 ± 0.038 | 0.879 ± 0.070 | 0.992 ± 0.028 | 0.957 ± 0.024 | 0.726 ± 0.147 | 0.580 ± 0.271 | 0.597 ± 0.325 | 0.652 ± 0.331 | 0.718 ± 0.170 |
| | NB | 0.827 ± 0.061 | 0.830 ± 0.060 | 0.818 ± 0.064 | 0.845 ± 0.072 | 0.827 ± 0.061 | 0.797 ± 0.144 | 0.690 ± 0.265 | 0.672 ± 0.303 | 0.784 ± 0.306 | 0.797 ± 0.161 |
| | AB | 0.960 ± 0.031 | 0.942 ± 0.048 | 0.939 ± 0.062 | 0.951 ± 0.075 | 0.958 ± 0.038 | 0.785 ± 0.142 | 0.642 ± 0.284 | 0.678 ± 0.333 | 0.680 ± 0.327 | 0.766 ± 0.173 |
| | GB | 0.930 ± 0.034 | 0.928 ± 0.036 | 0.959 ± 0.045 | 0.902 ± 0.063 | 0.930 ± 0.034 | 0.786 ± 0.139 | 0.623 ± 0.293 | 0.678 ± 0.342 | 0.646 ± 0.335 | 0.758 ± 0.168 |
| | RR | 0.848 ± 0.052 | 0.852 ± 0.052 | 0.826 ± 0.050 | 0.883 ± 0.070 | 0.848 ± 0.052 | 0.807 ± 0.129 | 0.710 ± 0.242 | 0.683 ± 0.280 | 0.812 ± 0.285 | 0.814 ± 0.145 |
| | NN | 0.894 ± 0.047 | 0.887 ± 0.054 | 0.942 ± 0.048 | 0.843 ± 0.085 | 0.894 ± 0.047 | 0.836 ± 0.132 | 0.709 ± 0.281 | 0.757 ± 0.318 | 0.732 ± 0.319 | 0.821 ± 0.154 |

Table 3. Results of 100 repetitions of 5-fold cross-validation. (Mean ± standard deviation). Logistic Regression (LR), Ridge Regression (RR), K-Nearest Neighbor (KNN), Naive Bayes (NB), Decision Tree (DT), Random Forest (RF), AdaBoost (AB), Gradient Boosting (GB) and Neural Network (NN).

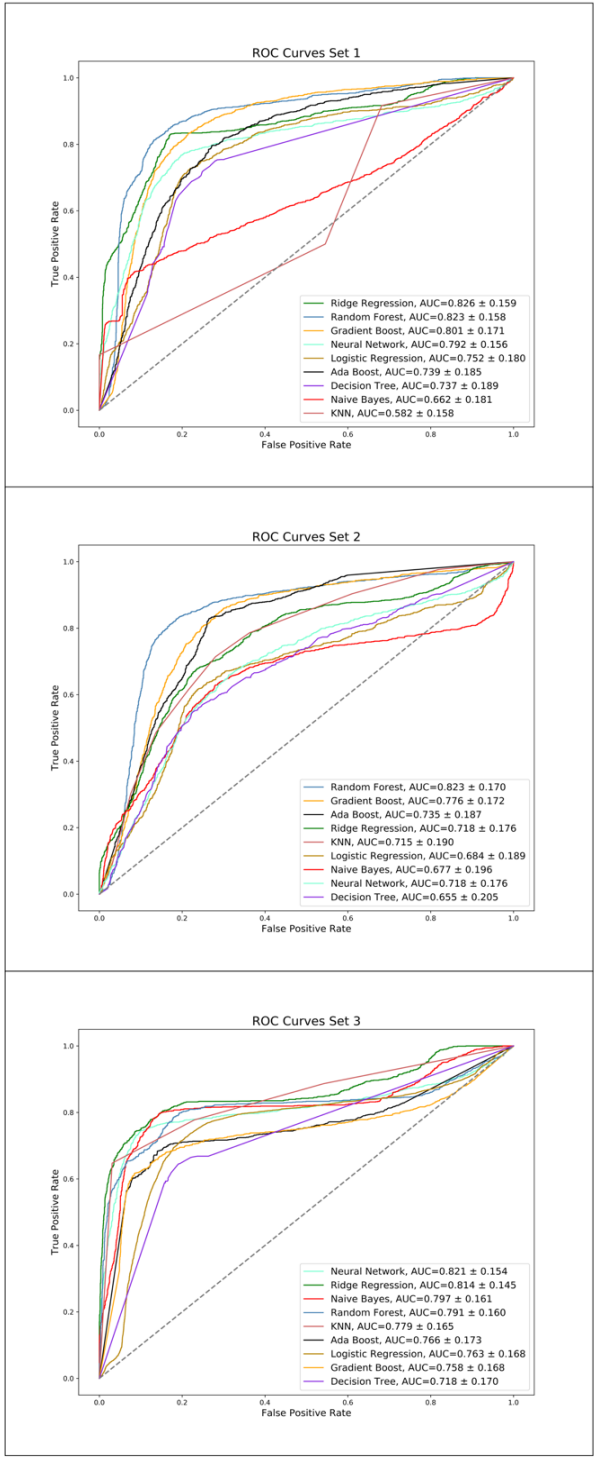

Fig. 3 ROC Curves of all models for each set of features.

## IV. Conclusion

We found a 35.2% prevalence of clinically relevant EGFR mutations, which is similar to results obtained by Zhang YL et al, meta-analysis that included 115.815 patients and found a pooled prevalence of 32.3% [8]. Regarding studies carried out in Chile, our results differ from those found by Gejman et al (prevalence of 21.7%) [9].

Detection of clinically relevant mutations has become one of the pillars of *Precision Medicine*, leading to great interest in non-invasive methodologies that can provide information in decision-making, allowing personalized diagnosis or providing relevant information on prognosis [10].

The underlying hypothesis of Radiomics is that advanced image analysis could capture information not currently used, specifically genomics and proteomics patterns that can be expressed in medical images [3].

Handcrafted features extraction involves manual segmentation of the region of interest (ROI) on medical imaging, and extraction of thousands of human-defined and curated quantitative features. This methodology has been well studied and has advantages in interpretability [11,12,13].

With this preprocessing and feature extraction methodology, we found imaging biomarkers that correlate with the presence of EGFR mutation with a predictive capacity in our models reached AUC of 0.826 (95% CI, 0.811 – 0.839), which agrees with other published works, such as the study by Zhang M et al. in which the radiomics model reached an AUC of 0.827 [4].

There was low concordance in the features selected by each method, except for skewness, which was repeated in the three sets with different filters. However, the performance in all sets was comparable. One of the important limitations in our work is the small number of patients, which may explain part of this variability.

The stability of radiomics features is a well-documented problem. There are multiple factors that can contribute to this variability, such as different equipment or acquisition protocols. In our case, the images of all patients were obtained in the same equipment, which provides homogeneous data. However, this may affect the reproducibility of the results when applied to patients with different imaging characteristics. The evaluation of differences between the features extracted by segmentations carried out by different operators is an element that can help prevent the bias induced by manual segmentation [14].

Deep Learning can solve some of the problems associated with variability in manual segmentation and perform automatic feature extraction without the bias generated by predefined handcrafted features. This translates into an increase in the predictive performance of Deep Learning models in prediction of EGFR in lung cancer patients demonstrated by other works [15,16,17]. One drawback of Deep Learning is loss of explainability of the models [13].

In summary, radiomics features analysis has the potential impact of predicting clinically relevant mutations in lung cancer patients through a non-invasive methodology without risk of complications and better availability. This allows a personalized choice of therapy, improving the prognosis and quality of life of patients.

### Acknowledgment

Thanks to Dra. Andrea Glasinovic and Dr. Hugo Figueroa for supporting research in the Radiology Department of Clínica Santa María.

### Conflict of Interest

The authors declare no conflict of interest.